# Skyrmion Lattice Order Controlled by Confinement Geometry

**Raphael Gruber[1], Jan Rothörl[1], Simon M. Fröhlich[1], Maarten A. Brems[1], Fabian Kammerbauer[1], Maria-Andromachi Syskaki[1,2], Elizabeth M. Jefremovas[1], Sachin Krishnia[1], Asle Sudbø[3], Peter Virnau[1], Mathias Kläui[1,3*]**

1. Institute of Physics, Johannes Gutenberg-Universität Mainz, Staudingerweg 7, 55128 Mainz, Germany.
2. Singulus Technologies AG, Hanauer Landstraße 103, 63796 Kahl am Main, Germany.
3. Center for Quantum Spintronics, Department of Physics, Norwegian University of Science and Technology, 7491 Trondheim, Norway.

*Email: klaeui@uni-mainz.de

This article may be downloaded for personal use only. Any other use requires prior permission of the author and AIP Publishing. This article appeared in *Appl. Phys. Lett.* **127**, 212403 (2025) and may be found at https://doi.org/10.1063/5.0299901.

## Abstract

Magnetic skyrmions forming two-dimensional (2D) lattices provide a versatile platform for investigating phase transitions predicted by Kosterlitz-Thouless-Halperin-Nelson-Young (KTHNY) theory. While 2D melting in skyrmion systems has been demonstrated, achieving controlled ordering in skyrmion lattices remains challenging due to pinning effects from a non-uniform energy landscape, which often results in polycrystalline structures. Skyrmions in thin films, however, offer thermal diffusion with high tunability and can be directly imaged via Kerr microscopy, enabling real-time observation of their dynamics. To regulate lattice order in such flexible systems, we introduce geometric confinements of varying shapes. Combining Kerr microscopy experiments with Thiele model simulations, we demonstrate that confinement geometry critically influences lattice order. Specifically, hexagonal confinements commensurate with the skyrmion lattice stabilize monodomain hexagonal ordering, while incommensurate geometries induce domain formation and reduce overall order. Understanding these boundary-driven effects is essential for advancing the study of 2D phase behavior and for the design of skyrmion-based spintronic applications, ranging from memory devices to unconventional computing architectures.

Magnetic skyrmions are topologically non-trivial chiral spin textures that exhibit quasi-particle behavior[1–3]. Their small size, stability, and dynamic properties make them highly promising for energy-efficient spintronic applications ranging from data storage[4] to sensing[5] and unconventional computing – where so far single skyrmions have been used[6–9]. For enhanced complexity however, arrangements of multiple skyrmions can be envisaged[10]. Beyond their technological significance for data processing, systems of dense skyrmion arrangements give rise to intriguing phenomena such as skyrmion drag[11]. And even denser lattice arrangements of skyrmions provide an ideal platform for exploring fundamental two-dimensional (2D) ordering phenomena[12–15] as quasi-long-range order (QLRO) can arise[12–16].

Skyrmion quasi-particles can exhibit thermally activated Brownian dynamics[17–19] and offer on-the-fly tunability of both their size and their diffusivity[18,20–22]. This versatility allows to investigate lattice phenomena beyond current observations and is a key advantage of skyrmions over other 2D systems like colloids or superconducting vortices. In particular, the tunability can be exploited to drive and observe 2D phase transitions[15] as described in Kosterlitz-Thouless-Halperin-Nelson-Young (KTHNY) theory[23–27]. These 2D phase transitions differ fundamentally from behavior in other dimensions, particularly from 3D. In particular, KTHNY theory describes the existence of a hexatic phase with only orientational QLRO between the solid phase (with translational QLRO) and the isotropic liquid (no QLRO)[25–27]. Eventually, these ordered lattice can even allow data transport by exploiting shock waves or the dynamics of topological lattice defects, which can move orders of magnitude faster than skyrmions[15,16]. Consequently, 2D phase transitions have attracted significant fundamental research interest for decades, both in theory and experiments[12,15,23–29].

In experimental skyrmion lattices, the key challenge in realizing QLRO is the underlying non-uniform energy landscape caused by material inhomogeneities[13–15,30]. The non-flat energy landscape describes a continuously varying potential with attractive as well as repulsive sites[18,31] – often commonly referred to as pinning effects. It causes quenched disorder, topological lattice defects and polycrystallinity, together breaking the QLRO of the lattice[13,30]. The local order is quantified by the orientational order parameter

$$\psi_6(\mathbf{r}_j) = \frac{1}{n}\sum_{k=1}^{n} e^{-i6\theta_{jk}} \quad (1)$$

for every skyrmion $j$ at position $\mathbf{r}_j$ and with $n$ nearest neighbors at positions $\mathbf{r}_k$ ($k$=1…$n$); where $\theta_{jk}$ denotes the angle between the horizontal axis (arbitrarily chosen) and the vector $\mathbf{r}_k$-$\mathbf{r}_j$ connecting the neighbor pair $j$ and $k$[25]. The Euler angle of the complex value of $\psi_6$ directly determines the local orientation $\alpha(\mathbf{r}_j)=\arg[\psi_6(\mathbf{r}_j)]/6$ of the lattice for every skyrmion. Regions of similar orientation $\alpha$ form a lattice domain. In a polycrystalline lattice, multiple lattice domains exist with different orientation of each domain. The domains are separated by boundaries at which $\alpha$ changes abruptly. These domain boundaries have been shown to be effectively pinned due to a non-flat energy landscape[30]. Thus, the lattice domains and their orientation appear pinned by the sample-specific energy landscape, imposing uncontrolled boundary conditions which are likely to be incommensurate with the ideal hexagonal skyrmion lattice.

To overcome this limitation, in this work, we artificially tune the boundary conditions by confining the skyrmion lattices inside different geometrical shapes. We find that commensurate shapes enhance the lattice order compared to an unconfined lattice, while the order is suppressed by incommensurate shapes. The results are consistent for Kerr microscopy experiments as well as Thiele model simulations[31–33].

We stabilize a polycrystalline skyrmion lattice close to room temperature (335 K) in a Ta(5 nm)/$Co_{20}Fe_{60}B_{20}$(0.9 nm)/Ta(0.07 nm)/MgO(2 nm)/Ta(5 nm) magnetic thin film multilayer stack with various confinement patterns (see supplementary material with Fig. S1 for details). The skyrmions are imaged in real-time (16 fps) and -space by using a commercially available Kerr microscope by *evico magnetics GmbH* using the polar magneto-optical Kerr effect. In Fig. 1a, we show a Kerr image of a skyrmion lattice in a hexagonal confinement. We use the *trackpy*[34] Python package to track the skyrmions and a Voronoi tessellation[35] to determine the lattice neighbors as well as the local order. After nucleation, the lattice evolves in time. In Fig. 1b, we color the local lattice orientation $\alpha$ for different snapshots of one video. While several small lattice domains are present in the beginning, the whole lattice aligns with the hexagonal confinement within minutes. The growth of the lattice domains is accelerated by an oscillating magnetic out-of-plane (OOP) field[20]. The hexagonal confinement is commensurate with the hexagonal skyrmion lattice structure and therefore allows stabilization of QLRO on this finite length scale[15].

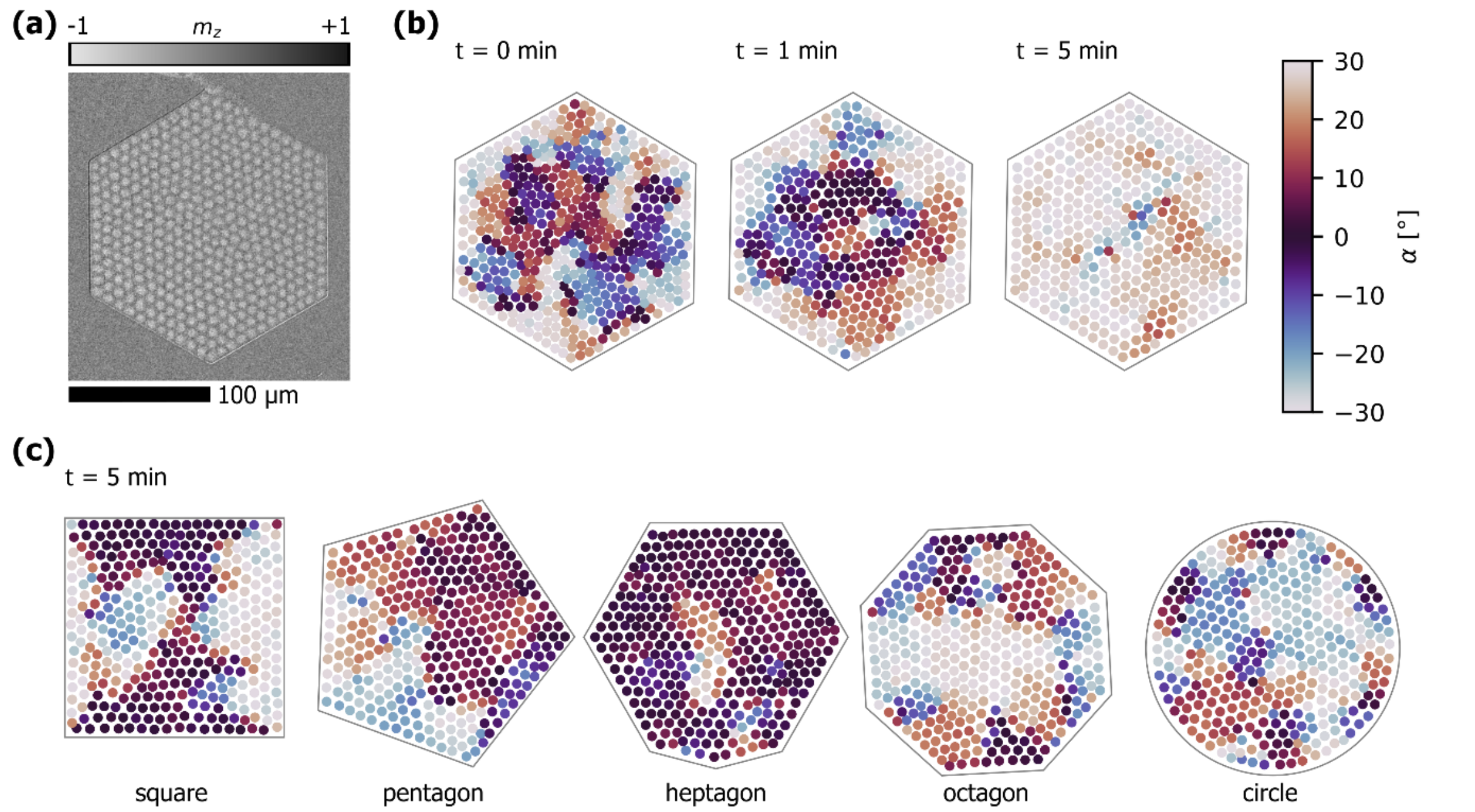

**Fig. 1: (a)** Polar Kerr microscopy image of a skyrmion lattice in a hexagonal confinement. **(b)** After nucleation, the skyrmion lattice arranges into hexagonal order on time scales of minutes within the commensurate hexagonal confinement, illustrated by the color-coded lattice orientation $\alpha$ per skyrmion for $t$=0,1,5 min after nucleation. Due to the six-fold symmetry of the hexagonal lattice, the color map of $\alpha$ is cyclic. **(c)** In different geometries, the incommensurate edges anchor different lattice domains, suppressing hexagonal order even after $t$=5 min.

As a comparison, in Fig. 1c, we present skyrmion lattices confined in different, incommensurate geometric shapes. All confinements are patterned on the same sample piece and have nominally the same area as the hexagon. Furthermore, all shapes are regular, except for the heptagon, which is a hexagon with an additional kink. In all geometries, the lattice locally aligns with the confinement edges and is therefore frustrated for the incommensurate shapes. Consequently, distinct lattice orientations are anchored at the boundary and enforce the occurrence of domain boundaries between each other[30]. While for the pentagon, every corner induces only a slight distortion, in total causing one domain boundary through the center, the many orientations around the octagon and the circle lead to much smaller areas of similar orientation. The distortions caused by the subtle irregularity of the heptagon can range from only slight misalignment of the outermost 2-3 skyrmion layers to even significant distortions far inside the lattice. In principle, a regular triangle would also be commensurate with the hexagonal lattice and therefore expected to enhance lattice order. However, edge effects may play a role in comparison to the hexagonal shape as the distances of every skyrmions to the edges are distributed differently.

To compare the lattice order in the different confinements quantitatively, we calculate the average local order $\Psi_6=\langle|\psi_6|\rangle$ as well as $\chi_6=|\langle\psi_6\rangle|$ in Fig. 2. For both parameters, higher values indicate better ordering. We furthermore compare the results to videos of a continuous sample of millimeter extension, where no confinements are patterned and around 3500 skyrmions are nucleated in the field of view. We find that the ordering is enhanced in the commensurate hexagon and suppressed for the incommensurate shapes. Thereby, $\chi_6$ reveals more drastic differences as it globally averages over $\alpha$, thus providing more long-range information but being significantly affected by changes of $\alpha$, that occur for instance at domain boundaries.

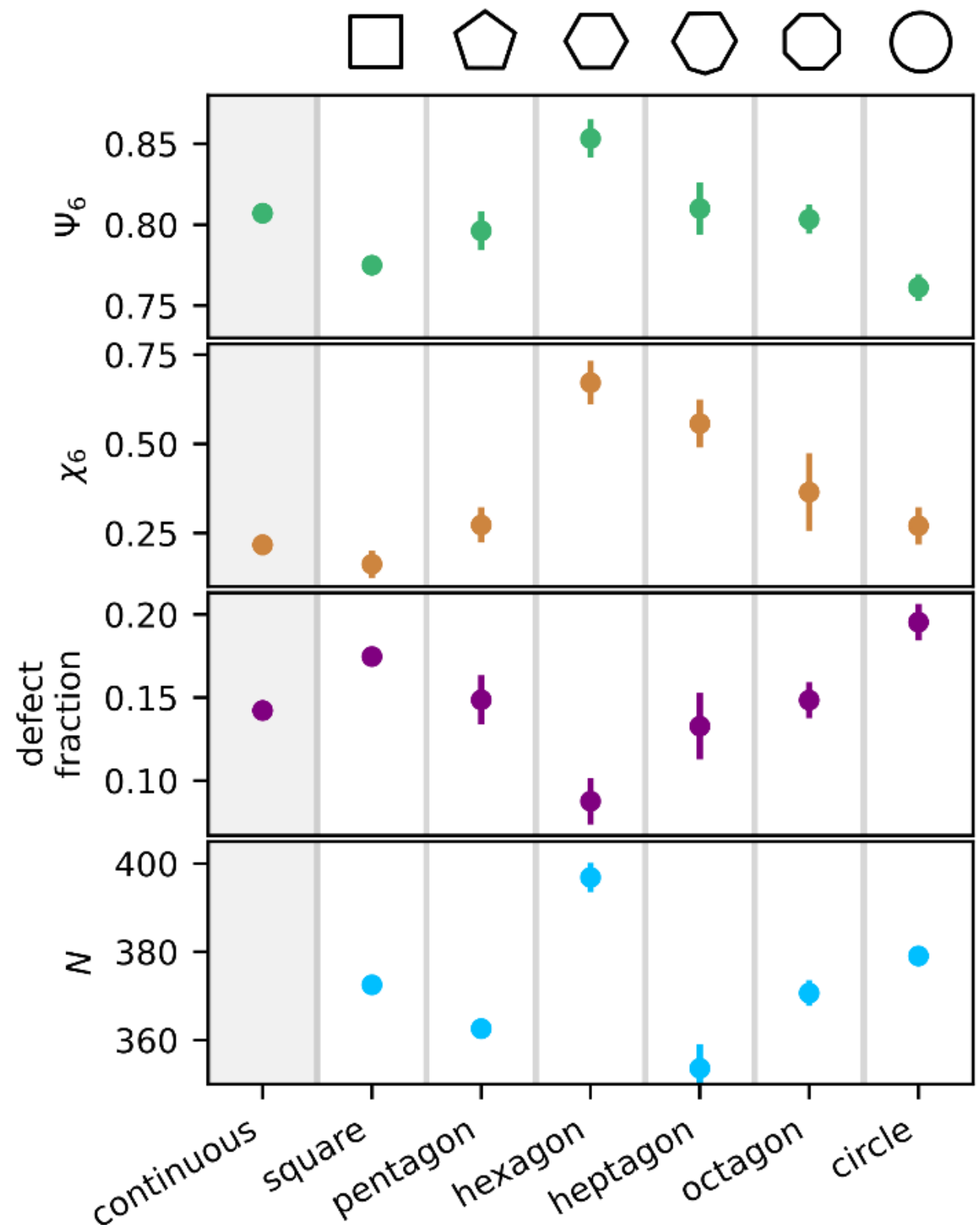

**Fig. 2:** Order parameters $\Psi_6$ and $\chi_6$, fraction of lattice defects and number of skyrmions $N$ compared for different confinement geometries and a lattice in an unpatterned, continuous film ($N$>3000) for reference. Data points are the average values from the last minute (4 min < $t$ < 5 min) of three independent videos. While the commensurate hexagonal confinement enhances the lattice order, incommensurate geometries suppress order with respect to the continuous case. From the mean values of the three different nucleations, we calculate the standard error of the mean as the error bar.

Correspondingly, the fraction of skyrmions being a topological lattice defect (i.e., $n\neq6$ lattice neighbors) increases for reduced ordering. The topological defects emerge as an intrinsic property of 2D lattices mediating their phase behavior[23–27], but also at domain boundaries or pinning sites. Even though a similar amount of $N$=510±20 skyrmions is created per nucleation (at $t$=0) in every geometry, we notice a clear difference in the annihilation rate for the different shapes: Skyrmion annihilation is less pronounced for the hexagon being commensurate with the hexagonal lattice, resulting in around $N$=395±5 skyrmions remaining at $t$=5 min; which is particularly close to the centered hexagonal number 401. In the incommensurate shapes however, the different lattice domains and domain boundaries cause space conflicts, which lead to skyrmion annihilations and result in only $N$=365±15 at $t$=5 min. In the continuous reference area, $N$=2620±15 skyrmions are left after 5 min, all of them are included to calculate the plotted order parameters. The time evolution of all shown parameters is presented in Fig. S2 in the supplementary material.

Our results demonstrate that confinement geometries play a critical role in stabilizing lattice order. By carefully designing the boundary conditions, however, the degree of order in skyrmion lattices can be effectively controlled – while the finite size itself has been shown only have a minor effect on the 2D phase behavior[15]. To understand the commensurability effect, we next perform Thiele model simulations[31–33] using confinements of different geometry.

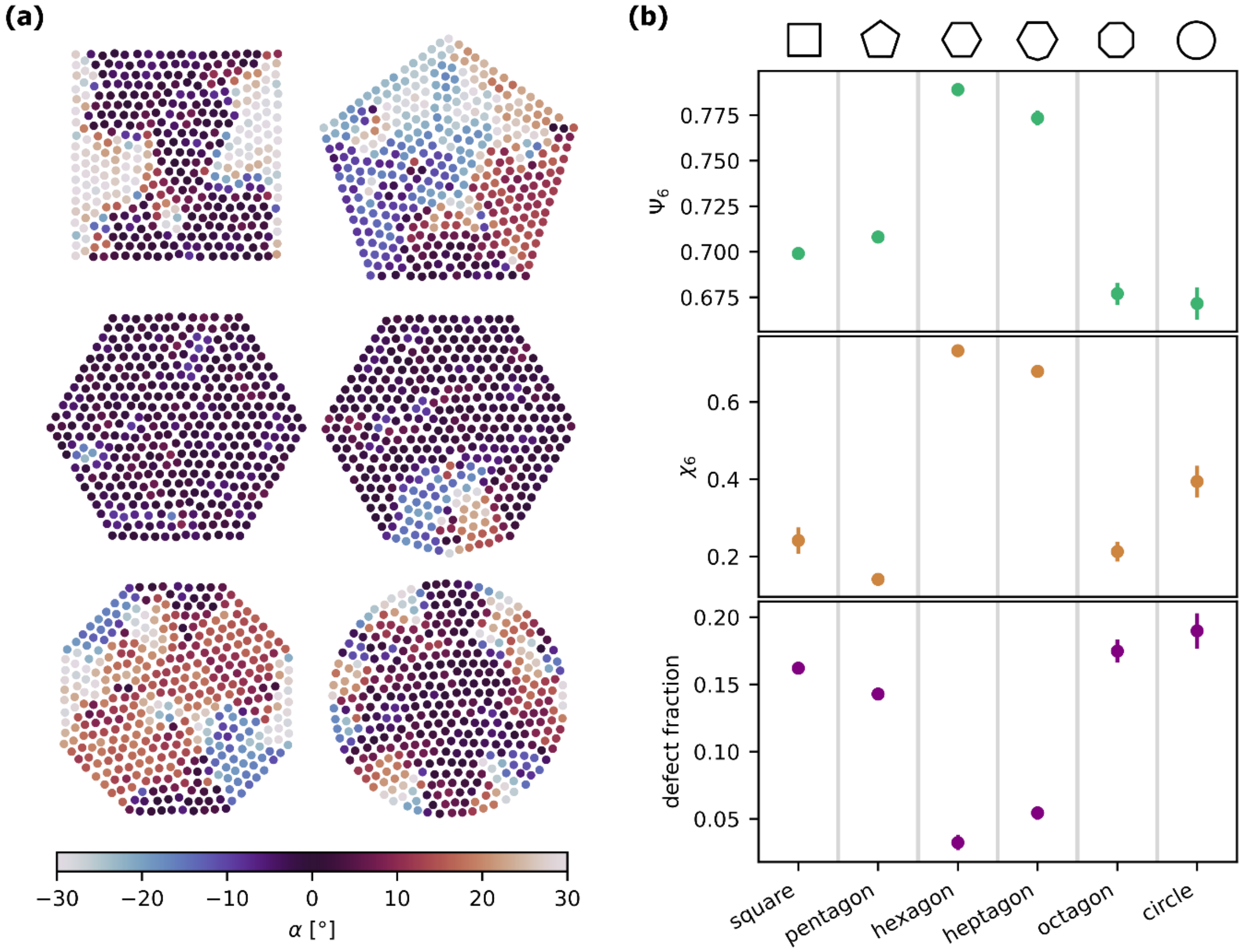

**Fig. 3: (a)** Local orientation $\alpha$ for simulated skyrmion lattices ($N$=400) in confinement, reproducing the alignment of lattice domains with the confinement edges and leading to multidomain states in incommensurate geometries. **(b)** The ordering parameters $\Psi_6$ and $\chi_6$ as well as the fraction of lattice defects behave similar to the experiment: the lattice order is significantly enhanced in the commensurate hexagon and the closely related heptagon. Data points are averages over 3×1000 snapshots, where the error bar is calculated as the standard error of the mean from the three independent data sets.

In the simulations, we employ a purely repulsive interaction potential of the form $r^{-8}$, which has been demonstrated to describe our experimental skyrmion system well[15,30,36] (see supplementary material with Fig. S3 for details). We find that the simulated particles align with the edges of the confinements as in the experiment and anchor differently oriented domains. We show examples of snapshots for every shape in Fig. 3a. The extent of the single domains varies between the 1000 snapshots of three different equilibration sequences of the system. However, the average ordering (reflected by $\Psi_6$, $\chi_6$, and the fraction of lattice defects) is consistent as presented in Fig. 3b. Especially, we reproduce the experimental result that the lattice order is suppressed in incommensurate geometric confinements. The error bars (calculated as standard error of the mean from the three equilibration sequences) are only visible for the incommensurate geometries as different lattice domains are forced to coexist but can have varying extent and distribution from snapshot to snapshot. For example, the distortion induced in the heptagon affects the 8 outermost skyrmion layers in Fig. 3a. However, in some equilibration sequences, it may only affect 2-3 layers but can – in analogy to the experiment – even reach the center of the geometry in other sequences, depending on the specific configuration. In contrast, the distribution of domains in the experiment is additionally influenced by the energy landscape which yields certain preferences for domain orientations and domain boundaries[30].

In conclusion, we demonstrate in both Kerr microscopy experiments and molecular dynamics simulations that the order of a confined magnetic skyrmion lattice can be tuned as it strongly depends on the confinement geometry. While a hexagonal confinement is commensurate with the hexagonal skyrmion lattice structure and stabilizes the lattice order, incommensurate geometries suppress order and differently oriented lattice domains form along the edges. In magnetic thin films, lattice domains are typically pinned in a similar way by the non-flat energy landscape[30]. Therefore, understanding those boundary effects is key to study 2D phase behavior with skyrmions on larger scales. Magnetic skyrmions are of special interest for investigating 2D phase behavior as their real-time accessibility in Kerr microscopy can allow one to investigate key open questions like the dynamics of topological defects[15] and their interaction potential – even in the presence of a Magnus force. Understanding and controlling the dynamics of densely packed skyrmions in a confined geometry[6,7,19] also plays a key role for realizing low-power non-conventional computing applications[5–7].

## Supplementary Material

See the supplementary material for a detailed description of the experimental and simulation setups.

## Acknowledgements

This work was funded by the Deutsche Forschungsgemeinschaft (DFG, German Research Foundation) - SPP 2137 (project #403502522), TRR 173/2 Spin+X (projects A01, A12 and B02). The authors acknowledge funding from TopDyn. This project has received funding from the European Research Council (ERC) under the European Union's Horizon 2020 research and innovation program (Grant No. 856538, project "3D MAGiC") and under the Marie Skłodowska-Curie grant agreements No. 860060 ("MagnEFi") and No. 101119608 ("TOPOCOM"). The authors gratefully acknowledge the computing time granted on the supercomputer MOGON II and III at Johannes Gutenberg University Mainz as part of NHR South-West. M.A.B. was supported by a doctoral scholarship of the Studienstiftung des deutschen Volkes. E.M.J. acknowledges the Alexander von Humboldt Postdoctoral Fellowship. A.S. and M.K acknowledge support from the Norwegian Research Council through Grant No. 262633, Center of Excellence on Quantum Spintronics (QuSpin). A. S. also acknowledges support from Norwegian Research Council through Grant No. 323766.

## Author Declaration Section

### Conflict of Interest

The authors have no conflicts to disclose.

### Author Contributions

R.G. performed the Kerr microscopy measurements and experimental data analysis. S.M.F., J.R. and M.A.B. conducted the MD simulations; R.G., J.R, S.M.F and M.A.B analyzed the simulation data. F.K. and M.A.S. optimized and fabricated the multilayer stack. R.G. prepared the manuscript with the help of J.R., M.A.B., E.M.J. and S.K.; A.S., P.V. and M.K. guided and supervised the work. All authors have commented on the manuscript.

## Data Availability Statement

The data that support the findings of this study are available on Zenodo under https://doi.org/10.5281/zenodo.17406257[37].

# Supplementary Material

## Magnetic Multilayer Material

The Ta(5 nm)/$Co_{20}Fe_{60}B_{20}$(0.9 nm)/Ta(0.07 nm)/MgO(2 nm)/Ta(5 nm) multilayer stack is deposited using DC/RF magnetron sputtering in a *Singulus Rotaris* system under a base pressure of $3\times10^{-8}$ mbar. The layer thickness is accurate to within 0.01 nm. The geometric confinements are patterned by electron beam lithography (EBL), followed by Argon ion etching. The continuous film reference measurements are performed on the same sample in a region of millimeter lateral dimension.

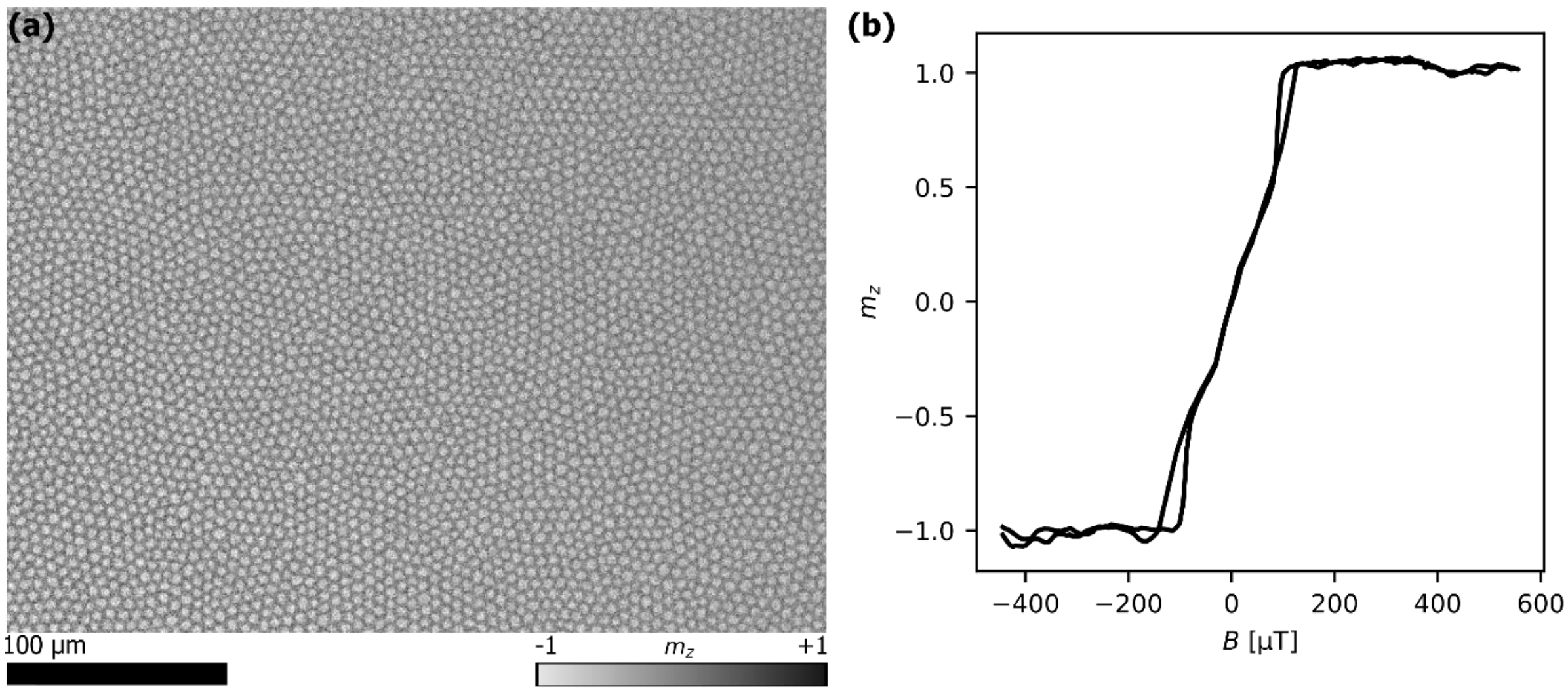

**Fig. S1. Skyrmion Lattice and Hysteresis Curve. (a)** Skyrmion lattice in the continuous film (millimeter extension) imaged by Kerr microscopy. The grayscale contrast denotes the out-of-plane (OOP) magnetization. **(b)** Relative OOP magnetization $m_z$ for an OOP field cycle measured by Kerr microscopy at 333 K. Note the small saturation field of less than 200 µT.

The interfacial Dzyaloshinskii-Moriya interaction (DMI)[1,2] arises primarily at the Ta/$Co_{20}Fe_{60}B_{20}$ interface, while the $Co_{20}Fe_{60}B_{20}$/MgO interface induces perpendicular magnetic anisotropy (PMA). We use a dusting layer of Ta(0.07) to balance DMI and PMA[3,4], thereby stabilizing skyrmions and optimizing the energy landscape for skyrmion lattice formation and dynamics. Skyrmions are stable in the continuous film as shown in Fig. S1a. For the patterned structures, the stack is physically removed to provide geometric confinement.

We provide the out-of-plane (OOP) hysteresis loop in Fig. S1b of the Supplementary Material to characterize the magnetic properties of the multilayer stack. Using spin-orbit torque-driven skyrmion motion and micromagnetic simulations, we confirm the non-trivial topology of the magnetic bubbles in our experiment[3,5–7].

Skyrmion Imaging and Order Analysis

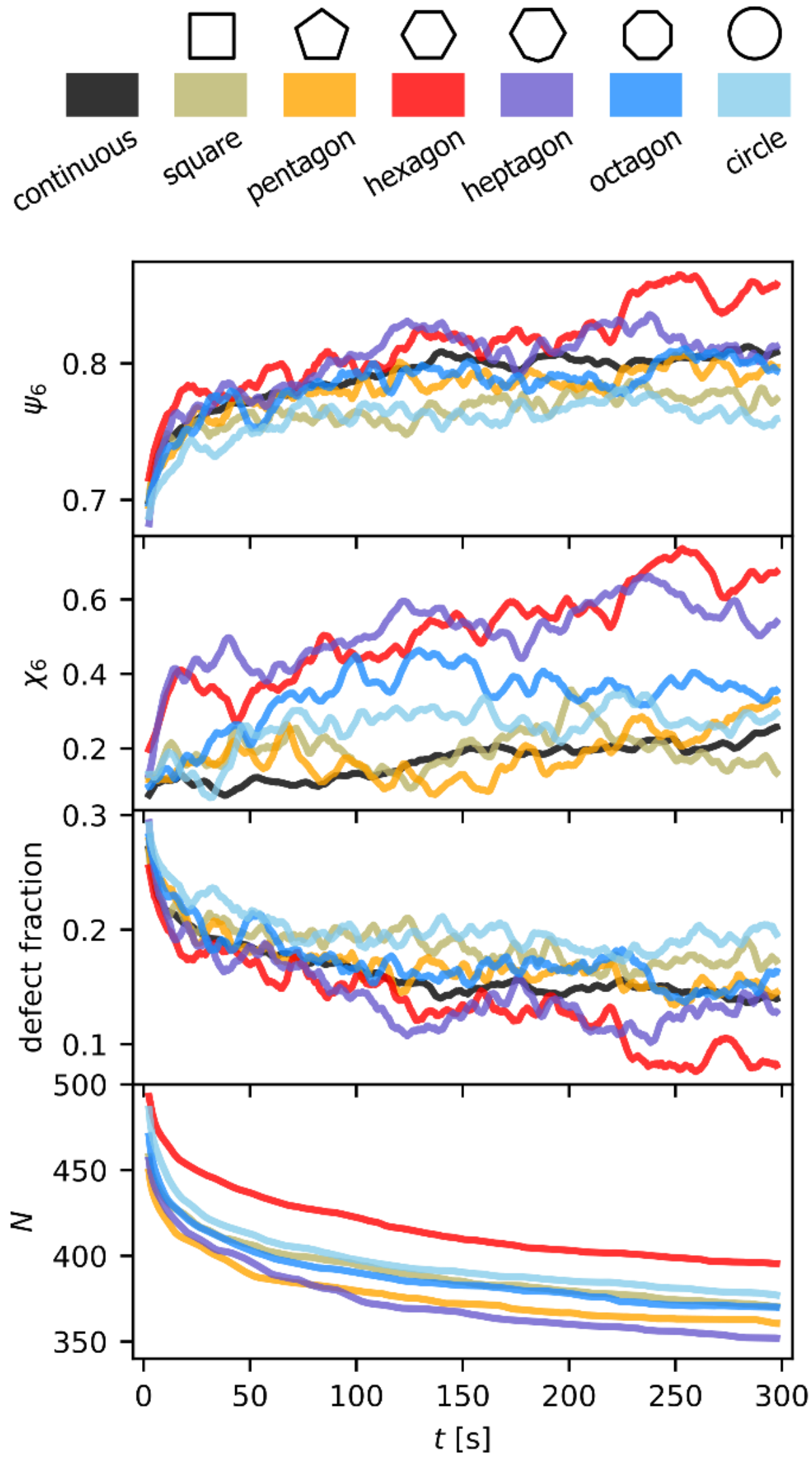

**Fig. S2**: Typical time evolution of the order parameters $\Psi_6$ and $\chi_6$, the fraction of lattice defects and the number of skyrmions *N* in the different confinements. For better visibility, only the rolling mean over 5 s is shown.

We establish magnetic contrast by magneto-optical Kerr effect in polar mode using a commercially available Kerr microscope manufactured by *evico magnetics GmbH* with a blue LED light source. We acquire Kerr images and videos (16 frames per second; 62.5 ms exposure time) with a CCD camera yielding gray-scale contrast at a field of view of 200×150 μm$^2$. We can control the magnetic field in in-plane (IP) and out-of-plane (OOP) direction separately by perpendicularly aligned electromagnetic coils. The OOP field coil is custom-made and allows for field control with a precision better than 1 μT. The fields are calibrated using a Hall probe and corrected for background fields by the offset of hysteresis loops[8,9]. A Peltier element on top of the coil allows for temperature control with a precision better than 0.1 K[3,6]. The thin film sample is placed directly on top of the Peltier element and temperature is monitored by a Pt100 sensor directly next to the sample. The whole setup is within a thermally stabilized flow box to improve stability of the operating conditions.

Skyrmions are nucleated by applying a large IP field pulse[5,10,11] while keeping the OOP field constant at its set value. The resulting skyrmions are a stable OOP domain state when the IP field is switched off again. We control the density and size of the skyrmions by the applied OOP field at a constant temperature[6,12,13]. We accelerate the formation of lattice order by reducing effective pinning with OOP field oscillations. We compare this formation process for several nucleations in videos capturing the 5 min after nucleation.

We use the 2D Gaussian kernel fitting within the *trackpy* Python package[14] to detect skyrmion positions in every frame. In the continuous reference area, the magnetic film extent of several millimeters exceeds the field of view significantly and is therefore considered as continuous, where we neglect boundary effects. We use a Voronoi tessellation[15] in every frame to extract nearest neighbor connections and the complex local order parameter $\psi_6$ for every skyrmion, where $\alpha=\arg(\psi_6)/6$ determines the local lattice orientation[16]. Every skyrmion with more or less than six nearest neighbors is a topological lattice defect[17,18].

## Thiele Model Simulations of Skyrmion Lattices

To simulate the thermal dynamics of skyrmions in a lattice configuration, we perform computer simulations. Since the experimental skyrmions are of micrometer size in a system spanning hundreds of micrometers, the skyrmions and especially their dynamics are not accessible in micromagnetic or atomistic simulations. Instead, we use a coarse-grained approach using the Thiele model[19]. The corresponding equation of motion reads[20,21]

$$-\gamma\mathbf{v} - G_{\mathrm{rel}}\gamma\mathbf{e}_z \times \mathbf{v} + \mathbf{F}_{\mathrm{therm}} + \mathbf{F}_{\mathrm{SkSk}}(\{\mathbf{r}\}) + \mathbf{F}_{\mathrm{SkBnd}}(\mathbf{r}) = 0 \qquad (S1)$$

with the set $\{\mathbf{r}\}$ of skyrmion positions $\mathbf{r}$, the skyrmion velocity $\mathbf{v}$, the total damping $\gamma$ (in the context of a Brownian Dynamics simulation, not the Gilbert damping) and the relative Magnus force strength $G_{\mathrm{rel}}$ (as tangent of the skyrmion Hall angle). We use $\gamma$=1 in simulation units. As $G_{\mathrm{rel}}$ is negligible in our system and furthermore only influences the lattice dynamics but not the static ordering, it is neglected here. The thermal Gaussian white noise $\mathbf{F}_{\mathrm{therm}}$ fulfils the fluctuation-dissipation theorem at a simulation unit temperature of $k_BT$=1. $\mathbf{F}_{\mathrm{SkSk}}$ and $\mathbf{F}_{\mathrm{SkBnd}}$ represent the repulsive skyrmion-skyrmion and skyrmion-boundary interaction. For the skyrmion-skyrmion interaction, a $V(r)=r^{-8}$ is used (cutoff distance of 1.8 simulation units[22]), which has been demonstrated to match the experimental system[8,20]. This skyrmion interaction potential has previously been determined in a very similar material stack by using Iterative Boltzmann Inversion (IBI)[20] without assuming any general form of the potential. The exact potential form used in this manuscript was however not determined from the conducted measurements presented here but in a less dense skyrmion liquid, as high density lattices generally lead to artefacts in the IBI[20]. In general, the skyrmion-skyrmion interaction in this system can be well described by both exponential and power-law forms since the potential within the cutoff radius is sufficiently steep. Choosing the power-law potential has the advantage that this form is widely studied in theory and allows comparison to previous works[8,20,22]. For the skyrmion-boundary interaction, we use a fully repulsive Lennard-Jones potential

$$V_{\mathrm{LJ}}(r) = 4\varepsilon\left[\left(\frac{\sigma}{r}\right)^{12} - \left(\frac{\sigma}{r}\right)^{6} + \frac{1}{4}\right] \qquad (S2)$$

with $r_{\mathrm{cut}}=2^{1/6}$ and $\varepsilon=\sigma=1$. We simulate systems of different skyrmion densities (see Fig. S3) by varying the spacing between skyrmions and set the density $\rho$ as the number of skyrmions per squared simulation unit length. With an Euler algorithm

$$\mathbf{r}(t+\Delta t) = \mathbf{r}(t) + \mathbf{v}(t)\Delta t \qquad (S3)$$

applying a time step of $\Delta t$=$10^{-4}$ in the *HOOMD-blue* software package[23], we determine the equation of motion.

The system is initialized with a square lattice and equilibrated for $10^6$ steps before running for $10^7$ steps with the trajectory saved every $10^4$ steps. Three independent equilibrations runs lead to 3×1000 saved position arrays (of all the 400 skyrmions) for every geometry.

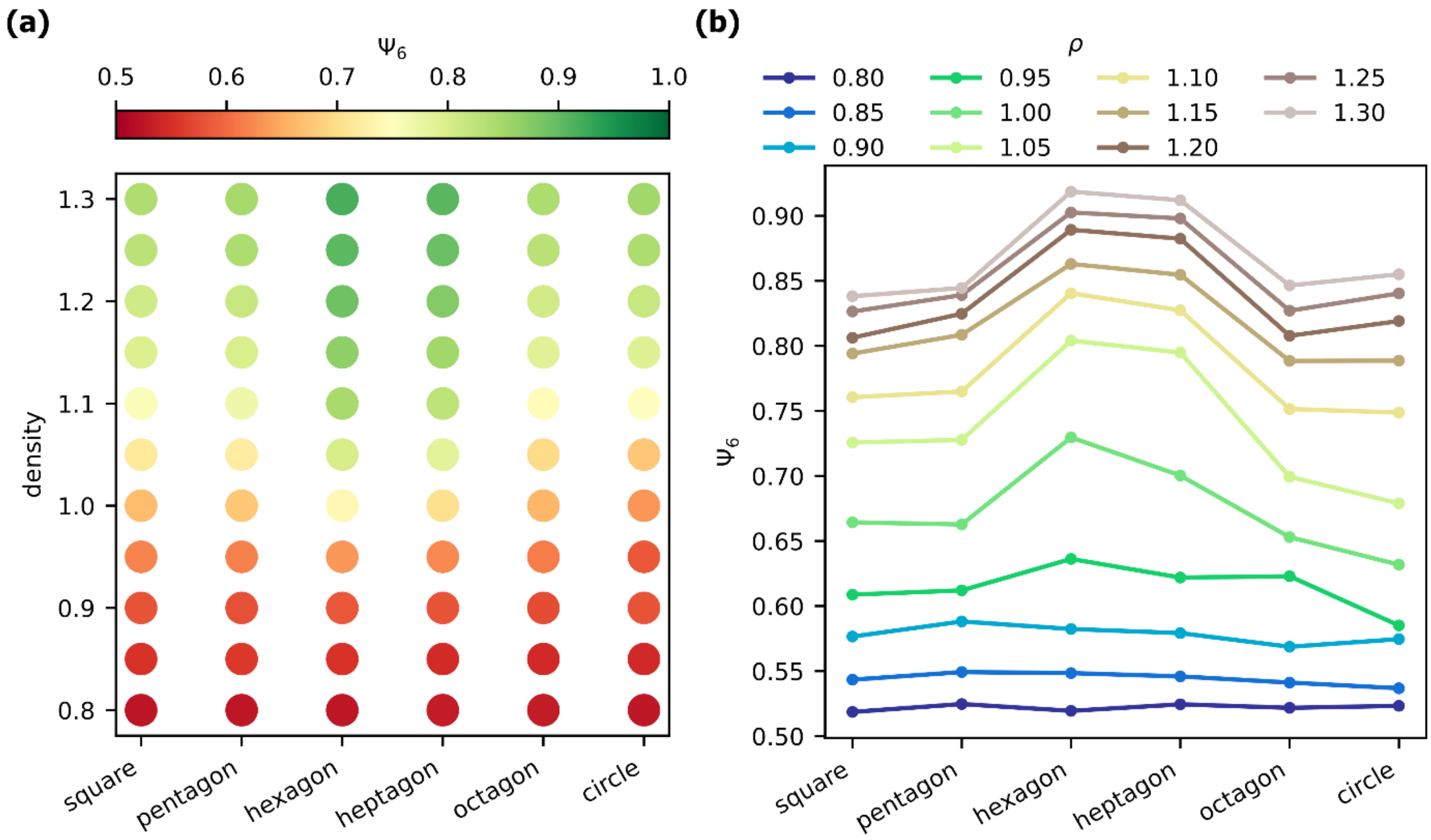

**Fig. S3.** Average local order $\Psi_6$ for Thiele model simulations of 400 skyrmions at different density $\rho$ and for different confinement geometries. **(a)** The resulting average value of $\Psi_6$ is presented as color-code for the different densities and confinement geometries. **(b)** The identical average values of $\Psi_6$ are now alternatively plotted for the different shapes (lines as guide to eye) where the color of the individual lines denotes the density $\rho$.

## References of the Supplementary Material